\documentclass[a4paper,11pt]{article}
\usepackage{graphicx}
\usepackage{xcolor}
\usepackage{amsmath,amssymb,amsfonts,authblk}%
\usepackage{amsthm}%
\usepackage{mathrsfs}%
\usepackage{booktabs}%
\usepackage{ctex}
\usepackage[colorlinks=true,bookmarks=true]{hyperref}
\usepackage{newtxtext}
\usepackage[backend=biber, style=phys, url=false]{biblatex}
\usepackage{pdfpages}
\usepackage[left=2.5cm,right=2.5cm,top=2.5cm,bottom=2.5cm]{geometry}

\usepackage{mathtools}

\DeclarePairedDelimiterX\braket[2]{\langle}{\rangle}{#1 \delimsize\vert #2}

\hypersetup{citecolor=blue}

\usepackage{setspace}
\begin{document}

\title{Manifold Optics}

\author[1,2,3]{Hongming Shen}
\author[2,3]{Wen Xiao}
\author[1]{Fei Fang Chung}
\author[1,2]{Huanyang Chen\footnote{Email: kenyon@xmu.edu.cn}}

\affil[1]{\normalsize{\textit{Department of Physics, Xiamen University Malaysia, Sepang 43900, Malaysia}}}

\affil[2]{\normalsize{\textit{Department of Physics, Xiamen University, Xiamen 361005, China}}}

\affil[3]{\normalsize{These authors contributed equally to this work.}}

\ctexset{today=old}
\date{}


\maketitle

\begin{abstract}
    Transformation optics establishes an equivalence relationship between gradient media and curved space, unveiling intrinsic geometric properties of gradient media. However, this approach based on curved spaces is concentrated on two-dimensional manifolds, namely curved surfaces. In this Letter, we establish an intrinsic connection between three-dimensional manifolds and three-dimensional gradient media in transformation optics by leveraging the Yamabe problem and Ricci scalar curvature, a measure of spatial curvature in manifolds. The invariance of the Ricci scalar under conformal mappings is proven. Our framework is validated through the analysis of representative conformal optical lenses.
  \end{abstract}

  A manifold is a topological space that locally resembles a Euclidean space $\mathbb{R}^{n}$~\cite{arbabOverviewManifoldIts1993, langDifferentialManifolds1985, leeIntroductionSmoothManifolds2012}. The local Euclidean property on a manifold allows the utilization of geometric and analytic techniques in higher dimensions. The curvature of a Riemannian manifold is characterized by the Ricci tensor, a fundamental object in differential geometry and general relativity. From the perspective of general relativity, it is the spacetime curvature, as is depicted by Ricci tensor, that induces gravitational effects in the Einstein field equations. In optics, transformation optics~\cite{leonhardtOpticalConformalMapping2006, pendryControllingElectromagneticFields2006, schurigCalculationMaterialProperties2006, chenTransformationOpticsMetamaterials2010} shares a close analogy with the equivalence principle in general relativity, providing a useful framework for understanding it by establishing a connection between the material parameters and curved space. It indicates that controlling light with gradient or anisotropic media can replicate the effects of light propagation in curved space, which implies the intrinsic geometric properties. For example, Maxwell's fisheye lens with refractive index profile $n=1/\left( 2+r^2 \right)$, being equivalent to 2-sphere, can be embedded into a three-dimensional (3D) flat space~\cite{xuLightRaysWaves2019}. And some other well-known absolute instruments perform perfect images possess curvatures in simple expressions, such as the Poincar\'e Disk with $n=1/\left( 2-r^2 \right)$ depicting the hyperbolic space, Luneburg lens with $n=\sqrt{2-r^2}$, Eaton lens with $n=\sqrt{2/r-1}$, etc. The intrinsic curvatures are usually characterized by the Ricci scalar~\cite{leonhardtGeometryLightScience2013} and Gaussian curvature~\cite{zhaoFlexuralWaveIllusion2023},
  $
   R=2K={2\left(\nabla n\right)^2}{n^{-4}}-{2{n^{-3}}\Delta n}=-{2}{n^{-2}} \Delta \log n,
  $
  where $R$ is the Ricci scalar, $K$ is the Gaussian curvature, $\Delta = \nabla ^2 = \partial ^2 / \partial x^2 + \partial ^2 / \partial y^2$ is the Laplacian, and $n=n\left(x,y\right)$ is the two-dimensional (2D) isotropic refractive index profile. The Ricci scalar is mathematically twice the Gaussian curvature, although they arise from completely different definitions. Additionally, the focusing mechanism has also been studied extensively in 2D curved spaces of constant curvature~\cite{batzLinearNonlinearOptics2008a, schultheissOpticsCurvedSpace2010}, where perfect imaging is observed.
  
  Nevertheless, despite this established fundamental connection, research on curvature-mediated light manipulation remains largely restricted to 2D gradient media (2-manifolds)~\cite{xuLightRaysWaves2019, xuObservationLightRays2020, zhaoFlexuralWaveIllusion2023}. Within this 2D paradigm, remarkable progress continues to be achieved. Of particular significance is the development of conformal transformation optics in two dimensions~\cite{liHidingCarpetNew2008, liUniversalMultimodeWaveguide2018, maApplicationInverseStrict2010, xuConformalTransformationOptics2015}, which has revolutionized the design schemes for photonic devices. Over the past two decades, this field has witnessed groundbreaking discoveries ranging from perfect imaging in constant positive-curvature spaces~\cite{schultheissOpticsCurvedSpace2010, schultheissLightCurvedTwodimensional2020}, geodesic framework~\cite{xuTheoryLightPropagation2021}, manipulation scheme~\cite{wangManipulatingCavityPhoton2022a}, etc. Non-Euclidean optical designs remain largely confined to two dimensions as well~\cite{geBilateralSymmetricNonEuclidean2024, leonhardtBroadbandInvisibilityNonEuclidean2009}. Given the captivating nature of 2D surfaces, the transition to 3D curved spaces promises to unveil fundamentally new physical phenomena and unconventional optical properties. However, research on modeling 3D optical systems using curved spaces remains scarce and challenging. This motivates a novel 3-manifold approach to extend beyond conventional 2D methods for modeling gradient media with intrinsic curvature. Such a framework would not only bridge this methodological gap but also significantly expand our capacity to engineer optical systems. New lens designs could emerge from curvature expressions of simple form. It's also worth noticing that recently, a novel method for designing multi-physical free-form metamaterials has been established based on the transformation theory~\cite{xuFreeformMultiphysicalMetamaterials2024}, further simplifying wave manipulation.

\begin{figure}[htbp]
      \centering
      \includegraphics[width=0.9\linewidth]{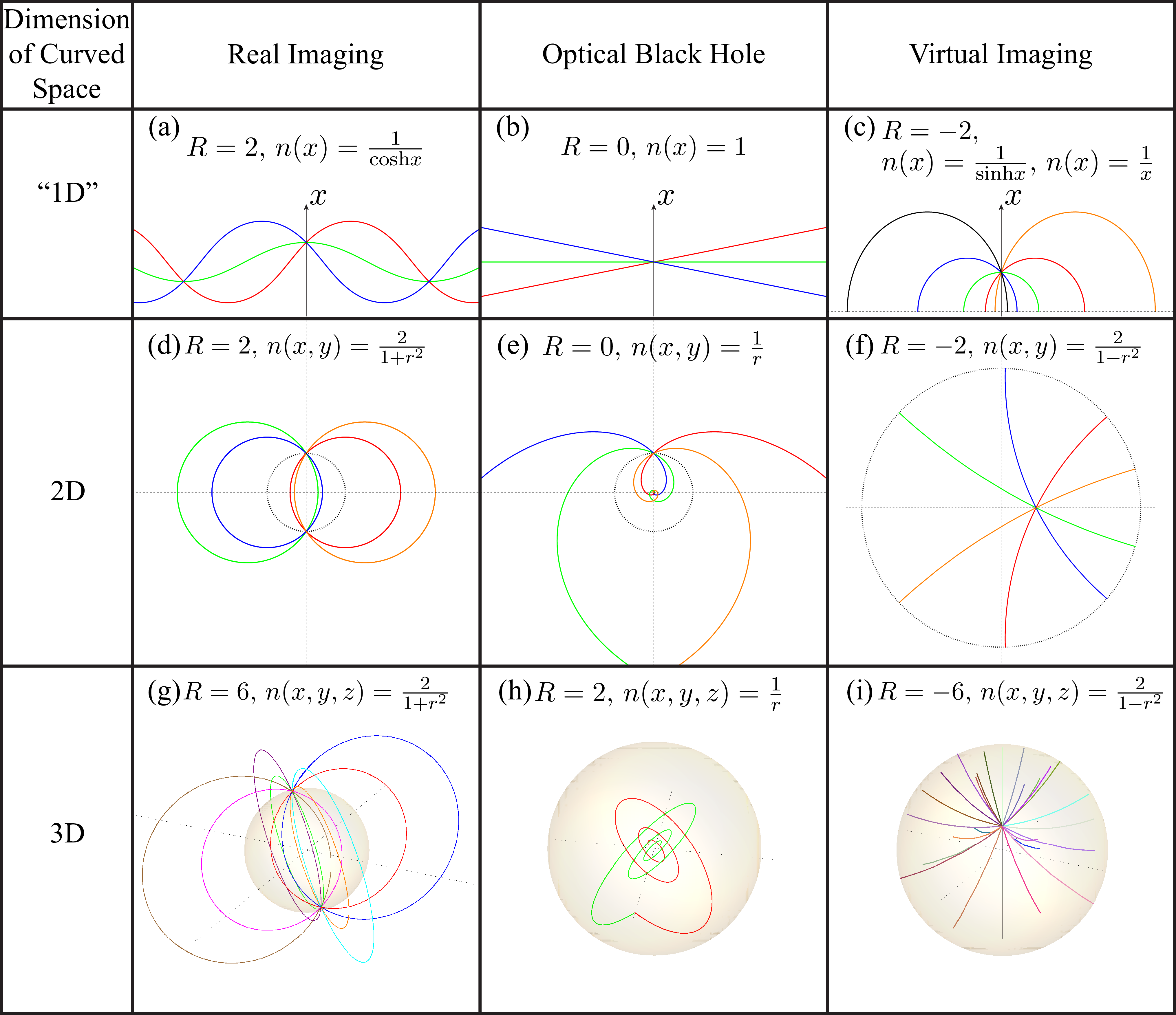}\caption{{Ray trajectories of an optical point source in gradient media from ``1D'' to 3D. (``1D'' indicates a profile varying along one coordinate only.)} (a) Real imaging with Mikaelian lens. (b) Flat space as an equivalence of an optical black hole with a singularity at infinity. (c) Virtual imaging at infinity with $n(x)=1/\text{sinh}x$ as well as $n(x)=1/x$. (d) Real imaging with 2D Maxwell's fisheye lens. (e) 2D optical black hole with its singularity centered at the origin. (f) Virtual imaging at the origin with Poincar\'e disk. (g) Real imaging with 3D Maxwell's fisheye lens. (h) 3D optical black hole with its singularity centered at the origin. (i) Virtual imaging at the origin with 3D Poincar\'e lens.}\label{fig1}
  \end{figure}

  In this Letter, we propose a scheme to connect gradient media with intrinsic curvatures by utilizing the Yamabe problem. Gradient refractive index media can be modeled as isotropic curved spaces, where intrinsic geometric properties strongly influence wave propagation. However, deriving these properties through differential geometry is usually computationally intensive. Our framework provides an easy way to obtain the essentials. Meanwhile, the invariance of curvature under conformal mapping is observed, thus demonstrating the power of manifold optics to determine imaging feature. Three types of imaging lenses of constant curvature, ranging from positive to negative, {Three types of imaging lenses with constant curvature, ranging from positive to negative and each exhibiting distinct imaging functionalities, are discussed} and illustrated in the columns of Fig.~\ref{fig1}, where conformal transformation connects the lenses in each type respectively.

  The Einstein field equations describe spacetime by inducing stress and energy with curvature in general relativity. Analogously, optics can also be governed solely by curvature in arbitrary dimensions. To achieve it, let's see how to derive the connection between a gradient medium and its Ricci scalar first. The traditional way to solve it involves lengthy derivation in differential geometry, {as in both relativity and optics.} The curved manifold of an isotropic gradient refractive index lens can be depicted by a metric $g_{ij}$, where the diagonal elements are $g_{ij} \delta _{ij} =n^2$ and off-diagonal elements are trivial,
  $
      \mathrm{d} s^2 = g_{ij} \mathrm{d} x^{i} \mathrm{d} x^{j}.
  $
  Here, we shall utilize the result from the Yamabe problem for a shortcut, which assumes a smooth positive function relating some manifolds to others of constant curvature. {Meanwhile, the shortcut restricts our media to be isotropic.} The Yamabe conjecture was solved in a manner that the explicit relation between the two manifolds ($g,g_0$) and the function $u$ is determined for arbitrary dimension $m \geq 3$~\cite{schoenConformalDeformationRiemannian1984},
  \begin{equation}\label{eq_02}
   -\Delta _{g_0} u + \frac{m - 2}{4(m - 1)} R_{g_0} u = \frac{m - 2}{4(m -1)}R_g u^{{\left(m + 2\right)}/{\left(m - 2\right)}},
  \end{equation}
  where $u$ is an arbitrary smooth positive function, $\Delta _{g_0}$ is the Laplace-Beltrami operator related to the metric tensor $g_0$, $R_g$ is the Ricci scalar related to the metric tensor $g$. $g_0$ is set as the Euclidean metric for convenience. And $g$ is set as the metric of the Riemannian manifold corresponding to the $m$-dimensional isotropic gradient medium, which means $g=g_{ij}=n^2 \mathbb{I}_{m}$ with $\mathbb{I}_{m}$ being $m$-dimensional identity matrix and $n$ being the isotropic refractive index profile. Furthermore, we can associate the conformal metric $g$ to the Euclidean metric by~\cite{chtiouiScalarCurvatureProblem2020},
  $
   g=u^{{4}/{\left(m -2\right)}}g_0 .
  $
  Recall that the metric for a curved manifold is defined as $g\equiv n^2 g_0$. Therefore, the smooth positive function $u$ can be related to the isotropic refractive index by comparison,
  \begin{equation}\label{eq_04}
      u^{{4}/{\left(m -2\right)}} = n^2.
  \end{equation}
  Notice that Euclidean leads to zero curvature, i.e., $R_{g_0}=0$. And so, $\Delta _{g_0}$ reduces to a Euclidean Laplacian. The subscript $g$ may be dropped for simplicity. Finally, we arrive at the generic form of the governing equation of manifold optics in $m$-dimensional gradient media by combining~\eqref{eq_02} and~\eqref{eq_04},
  \begin{equation}\label{eq_01}
   R=-{4\left(m -1\right)}/{\left(m -2\right)} {n^{-m /2 -1}} {\Delta n^{m / 2 -1}}, \, \forall m \geq 3.
  \end{equation}
  
  Since the study of gradient optics based on curvature in four dimensions or above still seem yet too far beyond current physical technology, we primarily focus on three dimensions. The governing equation of manifold optics in three dimensions is found by assigning dimension $m=3$ in~\eqref{eq_01},
  \begin{equation}\label{eq_06}
    R = -{8}{n^{-5/2}}\Delta \sqrt{n}.
  \end{equation}
  
  Before entering into the discussion of three typical types of optical instruments, we prove the invariance of curvature under 2D conformal mappings first. Consider a 2D conformal mapping that satisfies Cauchy-Riemann conditions from the $w$-space to $z$-space with coordinates $w=u+\mathrm{i}v$ and $z=x+\mathrm{i}y$. Consequently, the refractive index profiles in the two spaces are derived as~\cite{xuConformalTransformationOptics2015},
  $
   n_z = n_w \sqrt{\left( \frac{\partial u}{\partial x} \right)^2 + \left( \frac{\partial v}{\partial x} \right)^2 } = n_w \left| \frac{\mathrm{d}w}{\mathrm{d}z} \right|.
  $
  The Ricci scalars of the manifolds corresponding to the refractive index profiles remain unchanged as long as Cauchy-Riemann conditions are satisfied, i.e.,
  $
   R_z = -{2}{n_z^{-2}}\Delta \log n_z =-{2}{n_w^{-2}}\Delta ' \log n_w = R_w,
  $
  where $\Delta ' = {\partial ^2}/{\partial u^2} + {\partial ^2}/{\partial v^2}$ is the Laplacian in $w$-space. The full derivation is presented in Supplementary Note 1. The proof of curvature invariance under 3D conformal transformation is significantly more difficult due to the absence of Cauchy-Riemann condition alike constrains. Hence, the curvature invariance in three dimensions will be discussed case-by-case later.

  To start with, let's focus on a gradient medium with positive curvature, which is the 3D Maxwell's fisheye lens. It's refractive index profile is $n={2}/{\left(1+r^2\right)}$, where $r=\sqrt{x^2+y^2+z^2}$. From the governing equation of manifold optics~\eqref{eq_06}, the Ricci scalar of the 3D Maxwell's fisheye lens is derived as $R=6$. This is an optical instrument capable of perfect imaging, which can be seen from Fig.~\ref{fig1}(g). Analogous to the relationship between the 2D Maxwell's fisheye lens and its 3D spherical geodesic lens counterpart, the 3D Maxwell's fisheye lens can be regarded as a hypersphere embedded in four-dimensional (4D) space as for the geometric interpretation. In fact, the 3-sphere constitutes the 4D geodesic lens counterpart to the 3D Maxwell's fisheye lens, serving as an example of decreasing one dimension from a flat space to a curved space. To better demonstrate this dimensional decrease operation, coordinate systems are established in 4D space. Consider the 4D Cartesian coordinate system $\left\{x,y,z,w\right\}$. Then, the 3-sphere can be expressed as the collection of all points with a fixed distance from a point in four dimensions. Its center is set as the origin and radius as unity. In this way, the unit 3-sphere can be expressed as
  $
   x^2+y^2+z^2+w^2=1.
  $
  The spherical symmetry permits description using polar coordinates. Following Ref.~\cite{leonhardtGeometryLightScience2013}, a second polar angle, $\alpha$, is introduced to extend the polar coordinates from three dimensions to four dimensions, i.e., $\left\{r,\alpha , \theta , \phi\right\}$. Similarly, the conversion between Cartesian coordinates and polar coordinates can be written as
  \begin{equation}
      \begin{aligned}
          &x=r \sin \alpha \sin \theta \cos \phi, \quad
          y=r \sin \alpha \sin \theta \sin \phi, \\
          &z=r \sin \alpha \cos \theta, \quad
          w=r\cos \alpha,
      \end{aligned}
  \end{equation}
  where $r$ is the 4D radial distance, $\alpha \in \left[0,\pi\right]$ and $\theta \in \left[0,\pi\right]$ are the polar angles, and $\phi \in \left[0,2\pi\right)$ is the azimuthal angle. The unit 3-sphere is then expressed simply as $r=r_0\equiv 1$, which leaves us only three degrees of freedom, i.e., $\left\{\alpha, \theta, \phi\right\}$. To suppress a dimension, a curved 3D space with Cartesian coordinates $\left\{\tilde{x},\tilde{y},\tilde{z}\right\}$ shall be considered. The line element of the unit 3-sphere thus becomes
  \begin{equation}
      \begin{aligned}
          \mathrm{d}s^2 &= \mathrm{d}x^2 + \mathrm{d}y^2+ \mathrm{d}z^2+ \mathrm{d}w^2 \\
          &= r_0 ^2 \left[ \mathrm{d}\alpha ^2 +\sin ^2 \alpha \left(\mathrm{d} \theta ^2 +\sin ^2 \theta \mathrm{d} \phi ^2\right) \right]\\
          &= \tilde{n}^2 \left(\mathrm{d}\tilde{x}^2 +\mathrm{d}\tilde{y}^2+ \mathrm{d}\tilde{z}^2\right),
      \end{aligned}
  \end{equation}
  where $\tilde{n}=2/\left( 1+\tilde{x}^2+\tilde{y}^2+\tilde{z}^2 \right)$ is the refractive index of 3D Maxwell's fisheye lens. The last step is derived by applying the generalized stereographic projection, which is,
  $
   \tilde{x}={x}/\left( {1-w} \right),\,\tilde{y}={y}/\left( {1-w} \right),\,\tilde{z}={z}/\left( {1-w} \right).
  $
  Till this step, the 3-sphere is already expressed in a curved 3D space, i.e., the 3D Maxwell's fisheye lens. {During dimension suppression, the gradient media align perfectly with the curved surfaces embedded in a flat space of one higher dimension.}

  Conformal mapping in higher dimensions is also worth studying.  Consider a 3D conformal mapping, which is called the inverse mapping, to derive another gradient medium solution for $R=6$. The mapping is from virtual space $W^{i}=\{u,v,w\}$ to physical space $Z^{i}=\{x,y,z\}$, applied with following conformal mapping,
  \begin{equation}\label{eq_07}
   x=\frac{u}{u^2+v^2+w^2},\,
   y=\frac{v}{u^2+v^2+w^2},\,
   z=\frac{w}{u^2+v^2+w^2}.
  \end{equation}
  Under this mapping, the 3D Maxwell's fisheye lens centered at point $\left(a,b,c\right)$ in virtual space, which is
  $
   n_W = {2}/\{ {1+\left[(u-a)^2 + (v-b)^2 + (w-c)^2\right]} \},
  $
  can be transformed into the 3D Biased Maxwell's fisheye lens with a refractive index
  $
   n_Z ={2}/\left[ {\left( \pmb{r}\cdot \pmb{r}_c - 1 \right)^2 + \pmb{r}^2} \right],
  $
  where $\pmb{r}=x\hat{x}+y\hat{y}+z\hat{z}$ is the position vector and $\pmb{r}_c=a\hat{x}+b\hat{y}+c\hat{z}$ is the center. According to the governing equation of manifold optics~\eqref{eq_06}, one finds the curvature of the 3D Biased Maxwell's fisheye lens remains $R=6$. This is due to the curvature invariance of arbitrary gradient medium under inverse mapping in three dimensions. And the analytic proof is presented in Supplementary Note 2. The optical behavior of the 3D Maxwell's fisheye lens and the Biased one in terms of geometrical optics are both illustrated in Fig.~\ref{fig1}(g), as their ray trajectories are very similar, differing in translation only. The transparent sphere in the figure is a unit sphere for better visualization. Apart from the inverse conformal mapping, M\"obius mapping is also a conformal mapping but more generalized~\cite{leonhardtGeometryLightScience2013}. It is mixed with translation, stretching, and rotation. Due to the existence of two extra degree of freedoms defined in M\"obius mapping, it covers several other simple conformal mappings. The analytic proof of the invariance of curvature under 3D M\"obius transformation in three dimensions is presented in the Supplementary Note 3.

  3D Maxwell's fisheye lens with constant curvature at $R=6$ corresponds to perfect real imaging in 3D space. Meanwhile, a 2D Maxwell's fisheye lens with constant curvature at $R=2$ corresponds to perfect real imaging in 2D space, which is shown in Fig.~\ref{fig1}(d). {We also notice that spherical geometry is not necessary for perfect imaging. The 2D Generalized Maxwell's fisheye lens yields an example.} On the other hand, we define refractive indices dependent on only one coordinate as ``one-dimensional'' (``1D'') profiles for aesthetic reason, although the media are still in two dimensions. ``1D'' perfect real imaging can be achieved by exponent conformal mapping as $n=1/\cosh x$ (Mikaelian lens), which is shown in Supplementary Note 4. Undoubtedly, the gradient medium after transformation still possesses a constant Ricci scalar at $R=2$. And its perfect real imaging is demonstrated in Fig.~\ref{fig1}(a).
  
  It is empirically known that lenses with constant curvatures at $R=2$ in 2D space and $R=6$ in 3D space perform perfect imaging~\cite{schultheissOpticsCurvedSpace2010}, which corresponds to the ``Real Imaging'' column of Fig.~\ref{fig1}. However, we hereby propose that it may be a paradox. Although the 2D generalized Maxwell's fisheye lens whose refractive index is $n=2r^{1/M-1}/\left(1+r^{2/M}\right)$ possesses a constant curvature at $R=2/M^2$, it no longer remains valid when extended to three dimensions. The Ricci scalar of the 3D generalized Maxwell's fisheye lens is derived with the governing equation of manifold optics~\eqref{eq_06}. And its curvature can be expressed as
  $
   R=\left(1-M^2\right)\left(r^{2M}+r^{-2M}\right)/2+\left(1 + 5M^2\right),
  $
  which exhibits clear dependence on the radial distance $r$, except for Maxwell's fisheye lens ($M=\pm 1$) or the optical black hole ($M=0$). Nonetheless, the 3D generalized Maxwell's fisheye lens still performs at least one perfect imaging, which is also the self-imaging. Conversely, the lenses of constant curvature in three dimensions also don't imply perfect imaging, such as the 3D optical black hole we shall look into next.

  Consider the 3D optical black hole with a refractive index profile of $n=1/r$ as an example of the manifold with intrinsic curvature of $R=2$ calculated with the governing equation of manifold optics~\eqref{eq_06}. Its corresponding 3-manifold is a 3D hypercylinder, or 3-cylinder, embedded in 4D space. Likewise, the 3-cylinder is a 4D geodesic lens corresponding to the 3D optical black hole. Here, let us suppress the redundant dimension through geodesic mapping. We may consider the spherical coordinate system, $\{ \tilde{r}, \tilde{\theta}, \tilde{\phi} \}$, in curved 3D space, and the spherindrical coordinate system, $\{ r, \theta, \phi, w \}$, in flat 4D space. To clarify, the line element of the spherindrical coordinate system reads,
  \begin{equation}
      \begin{aligned}
              \mathrm{d}s^2 &= \mathrm{d}x^2 + \mathrm{d}y^2 + \mathrm{d}z^2 +\mathrm{d}w^2\\
              &= \mathrm{d}r^2 + r^2 \mathrm{d}\theta ^2 +r^2 \sin ^2 \theta \mathrm{d}\phi ^2 + \mathrm{d}w^2,
      \end{aligned}
  \end{equation}
  where $\{ x,y,z,w \}$ is the 4D Cartesian coordinate system. Then, the geodesic mapping can be expressed as
  $
  {\mathrm{d}h = \tilde{n} \mathrm{d}\tilde{r}},\, r=\tilde{n}\tilde{r},
  $
  where $\mathrm{d}h^2 = \mathrm{d}w^2 + \mathrm{d}r^2$. Meanwhile, the angles remain unchanged due to the rotational symmetry, i.e.,
  $
   \tilde{\theta} = \theta, \, \tilde{\phi} = \phi.
  $
  The 3-cylinder is the collection of all points at the same distance from the $w=0$ axis. Still, let the radius be unity for simplicity. The unit 3-cylinder thus can be expressed as
  $
   x^2+y^2+z^2=r^2=1.
  $
  Correspondingly, its line element can be written as
  $
   \mathrm{d}s^2=r_0^2 \mathrm{d}\theta ^2 + r_0^2 \sin \theta^2 \mathrm{d}\phi ^2 + \mathrm{d}w^2,
  $
  where $\mathrm{d}r=0$ is omitted and $r_0 =1$ is the unit radius. Hence, the certain geodesic mapping for 3-cylinder is
  $
   \mathrm{d}w=\tilde{n} \mathrm{d}\tilde{r}=\mathrm{d}\left(\log \tilde{r}\right),\, r_0=\tilde{n} \tilde{r},
  $
  which implies the curved space with a gradient refractive index profile
  $
   \tilde{n}=1/\tilde{r}=1/\sqrt{\tilde{x}^2 + \tilde{y}^2 + \tilde{z}^2}.
  $
  The optical behavior of the 3D optical black hole with constant curvature at $R=2$ is shown in Fig.~\ref{fig1}(h), where a point source is placed on the event horizon as well as the unit sphere. The 2D version is shown in Fig.~\ref{fig1}(e), which corresponds to a cylinder as a 2-manifold. Moreover, the ``1D'' optical black hole can be derived from the same exponent conformal mapping as well. The resulted medium is simply $n=1$ as is shown in Supplementary Note 4. Instead of converging to a singularity, the rays can be regarded as converging to infinity. Its ray trajectories are illustrated in Fig.~\ref{fig1}(b). {All the rays diverge to $r\rightarrow \infty$ in the ``1D'' optical black hole, which reveals the essence of divergence for higher dimensional optical black holes as well. The geometric rays in optical black holes do not form images and differ in their styles of divergence.}

  Lastly, let's examine some manifolds of constant negative curvature. The 3D Poincar\'e lens with refractive index $n={2}/{\left(1-r^2\right)}$~\cite{xiaoSitterSpaceGeneralized2022} falls into this category. The Ricci curvature is derived at $R=-6$ with~\eqref{eq_06}. Though the infinity refractive index on the unit sphere separated the whole space into two worlds, it still possesses a constant curvature of $R=6$ everywhere else. The optical behavior inside the unit sphere is shown in Fig.~\ref{fig1}(i). Ray trajectories intersect the unit sphere orthogonally, thus forming perfect 3D virtual images at the origin (for interior sources) or infinity (for exterior sources). Moreover, one can also apply the 3D inverse mapping~\eqref{eq_07}. Set the Poincar\'e lens centered at $\left(a,b,c\right)$ in virtual space
  $
   n_W = {2}/\{ {1-\left[(u-a)^2 + (v-b)^2 + (w-c)^2\right]} \}.
  $
  It can be transformed into the 3D Biased Poincar\'e lens with a refractive index profile
  $
   n_Z ={2}/\{ {\left( \pmb{r}\cdot \pmb{r}_c - 1 \right)^2 - \pmb{r}^2} \},
  $
  which only differs by a sign from the Maxwell's fisheye lens. The curvature remains $R=-6$, as shown by the governing equation of manifold optics~\eqref{eq_06}. The ray trajectories in this lens resemble the pre-transformation the Poincar\'e lens, with the space-partitioning sphere persisting, but radius and center are changed along with $\pmb{r}_c$.

  The 2D Poincar\'e Disk with constant intrinsic curvature of $R=-2$ also forms perfect 2D virtual images similarly, which is shown in Fig.~\ref{fig1}(f). Again, the gradient refractive lens for perfect ``1D'' virtual imaging can be derived with the same exponent conformal mapping as $n=1/\sinh x$ as is shown in Supplementary Note 4. The ray trajectories are demonstrated in Fig.~\ref{fig1}(c) accordingly. Likewise, the rays are attracted by an axis and intersect orthogonally, thus forming a ``1D'' virtual image at infinity.

  Furthermore, let us consider another 3-manifold with constant curvature at $R=-6$, which is $n(x,y,z)=1/x$. In this case, all the ray trajectories will be attracted by the singular plane, creating a virtual image at infinity. Its 2D and ``1D'' versions remain both unchanged, i.e., $n=1/x$. They possess constant curvature at $R=-6$ as well. The light propagation in its 2D and ``1D'' versions are both similar to what is shown in Fig.~\ref{fig1}(c).
  
  In conclusion, this is the first work proposing the intrinsic connection between 3-manifolds and 3D gradient optics. Such an equivalence provides a new approach to manipulating light propagation. The invariance of Ricci scalar under conformal mappings is shown. Our theoretical framework is further validated through analysis of several representative optical imaging lenses. The transformation from flat 4D spaces to curved 3D spaces facilitates investigation of 4D designs while providing new insights into 3D conformal mapping. {Extending the framework to anisotropic media remains an intriguing direction for future exploration. The study of chaos dynamics can also be extended into higher dimensions via our formalism \cite{xu2022light}.} Additionally, our scheme can be readily extended to other regimes by analogy, such as acoustic waves~\cite{chenAcousticCloakingTransformation2010, zhangBroadbandAcousticCloak2011, zhuSymmetricAcoustics2014, cummerControllingSoundAcoustic2016}, water waves~\cite{farhatBroadbandCylindricalAcoustic2008, hanWaterWavePolaritons2022, zhuControllingWaterWaves2024, zouBroadbandWaveguideCloak2019}, elastic waves~\cite{farhatUltrabroadbandElasticCloaking2009, farhatCloakingBendingWaves2009, miltonCloakingElasticityPhysical2006, stengerExperimentsElasticCloaking2012}{, offering potential for further applications, especially with experimental realizations supported by advances in artificial fabrication of 3D gradient materials \cite{xie2018acoustic, qu2024gradient},} which may be other powerful tools yet to explore.

\subsubsection*{Funding} This research was supported by the National Key Research and Development Program of China (2023YFA1407100); National Natural Science Foundation of China (12404371, 12361161667); China Postdoctoral Science Foundation (GZC20240906).
\subsubsection*{Disclosures} The authors declare no conflicts of interest.
\subsubsection*{Data availability} No data were generated or analyzed in the presented research.
\subsubsection*{Supplemental document} See Supplementary Notes for supporting content.

\printbibliography

\end{document}


\title{Manifold Optics: supplement}
\author{} 

\ctexset{today=old}
\date{}


\maketitle
\begin{abstract}
    This document provides supplementary information to ``Manifold Optics'', giving the details about related derivation.
\end{abstract}
\tableofcontents

\newpage

\section{Derivation of the invariance of Ricci curvature under 2D conformal transformation}
Consider a 2D conformal mapping that satisfies Cauchy-Riemann conditions from the $w$-space to $z$-space with coordinates $w=u+\mathrm{i}v$ and $z=x+\mathrm{i}y$, the refractive index are related by
\begin{equation}\label{eq03}
    n_z = n_w \sqrt{\left( \frac{\partial u}{\partial x} \right)^2 + \left( \frac{\partial v}{\partial x} \right)^2 } = n_w \left| \frac{\mathrm{d}w}{\mathrm{d}z} \right|.
\end{equation}
And the Ricci curvature of the gradient refractive index profile in $z$-space is
\begin{equation}\label{eq01}
    R_z = -\frac{2}{n_z^2}\Delta \log n_z.
\end{equation}
We may separate the Laplacian into two parts. Following the convention $z=x+\mathrm{i}y$ and $z^* = x-\mathrm{i}y$. Equivalently, $x=\left( z+z^* \right)/2$ and $y=\left( z-z^* \right)/\left( 2\mathrm{i} \right)$. By invoking the chain rule, partial derivatives of $x,y$ in terms of $z,z^*$ can be expressed as
\begin{equation}
    \begin{aligned}
        \frac{\partial}{\partial x} = \frac{\partial z}{\partial x} \frac{\partial}{\partial z} + \frac{\partial z^*}{\partial x} \frac{\partial}{\partial z^*} =& \frac{\partial}{\partial z} + \frac{\partial}{\partial z^*},\\
        \frac{\partial}{\partial y} = \frac{\partial z}{\partial y} \frac{\partial}{\partial z} + \frac{\partial z^*}{\partial y} \frac{\partial}{\partial z^*} =& \mathrm{i}\left( \frac{\partial}{\partial z} - \frac{\partial}{\partial z^*} \right)
    \end{aligned}
\end{equation}
Therefore, the Laplacian in $z$-space reads
\begin{equation}\label{eq02}
    \Delta = \left( \frac{\partial}{\partial x} \right)^2 + \left( \frac{\partial}{\partial y} \right)^2 = 4 \frac{\partial}{\partial z} \frac{\partial}{\partial z^*} =  4 \frac{\partial}{\partial z^*} \frac{\partial}{\partial z}.
\end{equation}
Likewise, the Laplacian in $w$-space would be in terms of $w,w^*$ with the same form which we shall use at the last step. Substituting~\eqref{eq02} back in to the Ricci curvature~\eqref{eq01},
\begin{equation}
    R_z = -\frac{8}{n_z^2} \frac{\partial}{\partial z}  \frac{\partial}{\partial z^*} \log n_z.
\end{equation}
Substituting~\eqref{eq03}, the Ricci curvature is expressed as
\begin{equation}
    \begin{aligned}
        R_z &= -\frac{8}{n_w^2 \left| \frac{\mathrm{d}w}{\mathrm{d}z} \right|^2} \left[ \frac{\partial}{\partial z}  \frac{\partial}{\partial z^*} \right] \log n_z\\
        &\text{(by chain rule)}\\
        &= -\frac{8}{n_w^2 \left| \frac{\mathrm{d}w}{\mathrm{d}z} \right|^2} \left[ \left( \frac{\partial}{\partial w} \frac{\partial w}{\partial z} \right)  \left( \frac{\partial}{\partial w^*} \frac{\partial w^*}{\partial z^*} \right) \right] \log n_z\\
        &\text{(by rearranging)}\\
        &= -\frac{8}{n_w^2 \left| \frac{\mathrm{d}w}{\mathrm{d}z} \right|^2} \left[ \left( \frac{\partial w}{\partial z} \frac{\partial w^*}{\partial z^*} \right)  \left( \frac{\partial}{\partial w} \frac{\partial}{\partial w^*}  \right) \right] \log n_z\\
        &\text{($w$ is dependent on $z$ only, and same for $w^*$ and $z^*$)}\\
        &= -\frac{8}{n_w^2 \left| \frac{\mathrm{d}w}{\mathrm{d}z} \right|^2} \left[ \left( \frac{\mathrm{d} w}{\mathrm{d} z} \frac{\mathrm{d} w^*}{\mathrm{d} z^*} \right)  \left( \frac{\partial}{\partial w} \frac{\partial}{\partial w^*}  \right) \right] \log n_z
    \end{aligned}
\end{equation}
We may utilize a very useful property in complex analysis,
\begin{equation}
    \left( \frac{\mathrm{d}w}{\mathrm{d}z} \right)^* = \frac{\mathrm{d}w^*}{\mathrm{d}z^*},
\end{equation}
which leads to
\begin{equation}
    \frac{\mathrm{d}w}{\mathrm{d}z} \frac{\mathrm{d}w^*}{\mathrm{d}z^*} = \left( \frac{\mathrm{d}w}{\mathrm{d}z} \right)^* \cdot \left( \frac{\mathrm{d}w}{\mathrm{d}z} \right) = \left| \frac{\mathrm{d}w}{\mathrm{d}z} \right|^2.
\end{equation}
Using this property, the curvature can be further reduced,
\begin{equation}\label{eq04}
    \begin{aligned}
        R_z &= -\frac{8}{n_w^2 \left| \frac{\mathrm{d}w}{\mathrm{d}z} \right|^2} \left[  \left| \frac{\mathrm{d}w}{\mathrm{d}z} \right|^2  \left( \frac{\partial}{\partial w} \frac{\partial}{\partial w^*}  \right) \right] \log n_z\\
        &\text{(Eliminating $\left| \frac{\mathrm{d}w}{\mathrm{d}z} \right|^2$)}\\
        &= -\frac{8}{n_w^2} \left( \frac{\partial}{\partial w} \frac{\partial}{\partial w^*}  \right) \log n_z\\
        &\text{(Express $n_z$ inside the logarithm in terms of $n_w$)}\\
        &= -\frac{8}{n_w^2} \left( \frac{\partial}{\partial w} \frac{\partial}{\partial w^*}  \right) \log \left( n_w \left| \frac{\mathrm{d}w}{\mathrm{d}z} \right| \right)\\
        &\text{(Using the properties of logarithm)}\\
        &=-\frac{8}{n_w^2} \left( \frac{\partial}{\partial w} \frac{\partial}{\partial w^*}  \right) \left[ \log \left( n_w \right) + \log \left( \left| \frac{\mathrm{d}w}{\mathrm{d}z} \right| \right) \right]\\
        &=-\frac{8}{n_w^2} \left( \frac{\partial}{\partial w} \frac{\partial}{\partial w^*}  \right) \left[ \log \left( n_w \right) + \log \left( \sqrt{\frac{\mathrm{d}w}{\mathrm{d}z} \frac{\mathrm{d}w^*}{\mathrm{d}z^*}} \right) \right]\\
        &=-\frac{8}{n_w^2} \left( \frac{\partial}{\partial w} \frac{\partial}{\partial w^*}  \right) \left[ \log \left( n_w \right) + \frac{1}{2} \log \left( \frac{\mathrm{d}w}{\mathrm{d}z} \frac{\mathrm{d}w^*}{\mathrm{d}z^*} \right) \right]\\
        &=-\frac{8}{n_w^2} \left( \frac{\partial}{\partial w} \frac{\partial}{\partial w^*} \right) \log \left( n_w  \right) -\frac{4}{n_w^2} \left( \frac{\partial}{\partial w} \frac{\partial}{\partial w^*} \right) \log \left( \frac{\mathrm{d}w}{\mathrm{d}z} \frac{\mathrm{d}w^*}{\mathrm{d}z^*} \right)\\
        &=-\frac{8}{n_w^2} \left( \frac{\partial}{\partial w} \frac{\partial}{\partial w^*} \right) \log \left( n_w  \right) -\frac{4}{n_w^2 } A,
    \end{aligned}
\end{equation}
where
\begin{equation}
    \begin{aligned}
        A &= \left( \frac{\partial}{\partial w} \frac{\partial}{\partial w^*} \right) \log \left( \frac{\mathrm{d}w}{\mathrm{d}z} \frac{\mathrm{d}w^*}{\mathrm{d}z^*} \right)\\
        &= \frac{\partial}{\partial w^*} \frac{\partial}{\partial w} \log \left( \frac{\mathrm{d}w}{\mathrm{d}z} \frac{\mathrm{d}w^*}{\mathrm{d}z^*} \right)\\
        &= \frac{\partial}{\partial w^*} \frac{\partial}{\partial w} \left( \log \frac{\mathrm{d}w}{\mathrm{d}z} + \log \frac{\mathrm{d}w^*}{\mathrm{d}z^*} \right) \\
        &= \frac{\partial}{\partial w} \left( \frac{\partial}{\partial w^*} \log \frac{\mathrm{d}w}{\mathrm{d}z} \right) + \frac{\partial}{\partial w^*} \left( \frac{\partial}{\partial w} \log \frac{\mathrm{d}w^*}{\mathrm{d}z^*} \right)\\
        &= \frac{\partial}{\partial w} \left( \frac{\partial z^*}{\partial w^*} \frac{\partial}{\partial z^*} \log \frac{\mathrm{d}w}{\mathrm{d}z} \right) + \frac{\partial}{\partial w^*} \left( \frac{\partial z}{\partial w} \frac{\partial}{\partial z} \log \frac{\mathrm{d}w^*}{\mathrm{d}z^*} \right)\\
        &= \frac{\partial}{\partial w} \frac{\partial z^*}{\partial w^*} \left( \frac{\partial}{\partial z^*} \log \frac{\mathrm{d}w}{\mathrm{d}z} \right) + \frac{\partial}{\partial w^*} \frac{\partial z}{\partial w} \left( \frac{\partial}{\partial z} \log \frac{\mathrm{d}w^*}{\mathrm{d}z^*} \right).
    \end{aligned}
\end{equation}
Since $w$ is a function dependent on $z$ only, indicating that $w$ is independent of $z^*$, $\log \frac{\mathrm{d}w}{\mathrm{d}z}$ will be only dependent on $z$. Taking a partial derivative of such a function independent on $z^*$ with respect to $z^*$ leads to zero. Similar argument explains the other term, $\frac{\partial}{\partial z} \log \frac{\mathrm{d}w^*}{\mathrm{d}z^*}$, goes to zero as well. Hence,
\begin{equation}
    \begin{aligned}
        A &= \frac{\partial}{\partial w} \frac{\partial z^*}{\partial w^*} \cdot 0 + \frac{\partial}{\partial w^*} \frac{\partial z}{\partial w} \cdot 0 = 0
    \end{aligned}
\end{equation}
Returning to~\eqref{eq04}, the curvature can be further reduced to
\begin{equation}
    \begin{aligned}
        R_z &=-\frac{8}{n_w^2} \left( \frac{\partial}{\partial w} \frac{\partial}{\partial w^*} \right) \log \left( n_w  \right)\\
        &=-\frac{2}{n_w^2} \left( 4 \frac{\partial}{\partial w^*} \frac{\partial}{\partial w} \right) \log \left( n_w  \right)\\
        &=-\frac{2}{n_w^2} \left( \frac{\partial ^2}{\partial u^2} + \frac{\partial ^2}{\partial v^2}  \right) \log \left( n_w  \right)\\
        &=-\frac{2}{n_w^2}\Delta ' \log n_w = R_w,
    \end{aligned}
\end{equation}
where $\Delta ' = \frac{\partial ^2}{\partial u^2} + \frac{\partial ^2}{\partial v^2}$ is the Laplacian in $w$-space.


\section{Proof of the invariance of Ricci curvature under 3D inverse mapping}
In this section, we prove the invariance of Ricci curvature of an arbitrary isotropic gradient refractive index profile under the 3D inverse mapping, which is a type of 3D conformal transformation. We consider the inverse mapping from virtual space $W^{i}=\left\{u,v,w\right\}$ to physical space $Z^{i}=\left\{x,y,z\right\}$,
\begin{equation}\label{eq05}
        x=\frac{u}{u^2+v^2+w^2},\quad
        y=\frac{v}{u^2+v^2+w^2},\quad
        z=\frac{w}{u^2+v^2+w^2}.
\end{equation}
Equivalently, the coordinates for virtual space can be solved in terms of the physical coordinates as
\begin{equation}\label{eq06}
        u=\frac{x}{x^2+y^2+z^2},\quad
        v=\frac{y}{x^2+y^2+z^2},\quad
        w=\frac{z}{x^2+y^2+z^2}.
\end{equation}
To apply coordinate transformation, we consider the isotropic metric tensor of $W$ space as 
\begin{equation}\label{eq10}
    g_{ij} = \mathrm{diag}\left\{{n_W}^2,{n_W}^2,{n_W}^2\right\}.
\end{equation}
To simplify the coordinate transformation, we let the permittivity and the permeability tensors before equal, i.e.,
\begin{equation}
    \epsilon_W ^{ij} = \mu_W ^{ij} = \sqrt{g}g^{ij},
\end{equation}
where $g^{ij}=\left(g_{ij}\right)^{-1}$ is the contravariant tensor and $g=\mathrm{det}\left(g_{ij}\right)$.
In this way, we obtain the material tensors relate to the refractive index by
\begin{equation}
    \epsilon_W ^{ij} = \mu_W ^{ij} = \mathrm{diag}\left\{{n_W},{n_W},{n_W}\right\},
\end{equation}
where $n_W$ is the gradient refractive index profile in virtual space. According to transformation optics, the permittivity tensor and the permeability tensors after transformation are given by
\begin{equation}
    \epsilon_Z^{ij} = \frac{A\epsilon_W^{ij} A^{\mathrm{T}}}{\left| \mathrm{det}A \right|},\quad \mu_Z^{ij} =\frac{A\mu_W^{ij} A^{\mathrm{T}}}{\left| \mathrm{det}A \right|},
\end{equation}
where
\begin{equation}
    A={A^Z}_W = \frac{\partial Z}{\partial W} =
    \begin{bmatrix}
        \frac{\partial x}{\partial u} & \frac{\partial x}{\partial v} & \frac{\partial x}{\partial w} \\
        \frac{\partial y}{\partial u} & \frac{\partial y}{\partial v} & \frac{\partial y}{\partial w} \\
        \frac{\partial z}{\partial u} & \frac{\partial z}{\partial v} & \frac{\partial z}{\partial w}
    \end{bmatrix}
\end{equation}
is the Jacobian transformation matrix. The material tensors after transformation are still equal and diagonal, i.e.,
\begin{equation}\label{eq11}
    \epsilon_Z ^{ij} = \mu_Z ^{ij} = \mathrm{diag}\left\{{n_Z},{n_Z},{n_Z}\right\},
\end{equation}
where $n_Z$ is the isotropic gradient refractive index profile after transformation (in physical space). For inverse mapping~\eqref{eq05}, the relation between two refractive index profiles are
\begin{equation}
    n_{Z} = n_{W} \left( u^2+v^2+w^2 \right).
\end{equation}
Therefore, the Ricci curvature for physical space is
\begin{equation}\label{eq07}
    \begin{aligned}
        R_{Z} &= -\frac{8}{n_{Z}^{5/2}}\left(\partial_{x}^2+\partial_{y}^2+\partial_{z}^2\right)\sqrt{n_Z}\\
        &=-\frac{8}{n_{W}^{5/2}\left( u^2+v^2+w^2 \right)^{5/2}}\left(\partial_{x}^2+\partial_{y}^2+\partial_{z}^2\right)\sqrt{n_W \left( u^2+v^2+w^2 \right)}\\
    \end{aligned}
\end{equation}
The derivatives in terms of physical coordinates can be expressed in terms of the virtual coordinates with the help of chain rule by regarding $\left\{ u,v,w \right\}$ as functions of $\left\{ x,y,z \right\}$. Additionally,~\eqref{eq06} provides enough information to make the terms produced in chain rule explicit, which are in fact the elements of the Jacobian $\frac{\partial W}{\partial Z}$. And the Ricci scalar in terms of virtual-coordinate-derivatives reduces to
\begin{equation}\label{eq14}
    \begin{aligned}
        R_{Z}&=\frac{2}{n_W^4} \left[ \left( \nabla n_W \right)^\mathrm{T}\cdot \left( \nabla n_W \right) -2n_W \Delta n_W \right]\\
        &= -\frac{8}{n_W ^{5/2}}\Delta \sqrt{n_W}=R_W,
    \end{aligned}
\end{equation}
where $\nabla = \partial _{u}\hat{u} + \partial _{v}\hat{v} + \partial _{w}\hat{w}$ is the gradient in terms of virtual coordinates, $\Delta = \frac{\partial ^2}{\partial u^2} + \frac{\partial ^2}{\partial v^2} + \frac{\partial ^2}{\partial w^2}$ is the Laplacian in virtual space, and $R_W$ is the Ricci curvature in virtual space. Hence, the invariance of Ricci curvature under 3D inverse mapping is proven.

\section{Proof of the invariance of Ricci curvature under 3D M\"obius mapping}
Just like the previous section, we provide the proof of invariance of Ricci curvature under another 3D conformal mapping. We start with the spatial M\"obius mapping,
\begin{equation}
    \pmb{r}' = \frac{ \pmb{r} \left( 1+r_0^2/a^2 \right) - \pmb{r_0} \left( 1+2\pmb{r_0}\cdot \pmb{r}/a^2 - r^2/a^2 \right) }{ 1+2\pmb{r}\cdot \pmb{r_0}/a^2 + r^2r_0^2/a^4 },
\end{equation}
where $\pmb{r} = u\hat{u} + v\hat{v} + w\hat{w}$ is the spatial coordinates in virtual space, $\pmb{r_0} = a\mathrm{tan}\gamma \hat{u} \equiv at \hat{u}$ depicts the M\"obius rotation angle of the transformation with $a$ and $\gamma$ being arbitrary parameters, and $\pmb{r'} = x\hat{x} + y\hat{y} + z\hat{z}$ is the spatial coordinates in physical space. Separating the vector transformation form into Cartesian components, we derive the 3D M\"obius mapping as below,
\begin{equation}\label{eq08}
    \begin{aligned}
        x &= \frac{ a\left[ t\left( u^2+v^2+w^2-a^2 \right) + \left( 1 - t^2 \right)au \right] }{ t^2 \left( u^2+v^2+w^2 \right) + 2atu + a^2 },\\
        y &= \frac{ a^2\left( t^2 + 1 \right)v }{ t^2 \left( u^2+v^2+w^2 \right) + 2atu + a^2 },\\
        z &= \frac{ a^2\left( t^2 + 1 \right)w }{ t^2 \left( u^2+v^2+w^2 \right) + 2atu + a^2 }.\\
    \end{aligned}
\end{equation}
Likewise, the virtual coordinates can be solved to be expressed in terms of the physical ones,
\begin{equation}\label{eq09}
    \begin{aligned}
        u &= \frac{ a\left[ -t\left( x^2+y^2+z^2-a^2 \right) + \left( 1 - t^2 \right)ax \right] }{ t^2 \left( x^2+y^2+z^2 \right) + 2atx + a^2 },\\
        v &= \frac{ a^2\left( t^2 + 1 \right)y }{ t^2 \left( x^2+y^2+z^2 \right) + 2atx + a^2 },\\
        w &= \frac{ a^2\left( t^2 + 1 \right)z }{ t^2 \left( x^2+y^2+z^2 \right) + 2atx + a^2 }.\\
    \end{aligned}
\end{equation}

Again, coordinate transformation is carried out with similar procedure explained from~\eqref{eq10} to~\eqref{eq11}, and the resultant relation between the two refractive index profiles are
\begin{equation}\label{eq12}
    n_Z = n_W \cdot \frac{ \mathrm{abs} \left[ a^2 + 2atu + t^2 \left( u^2+v^2+w^2 \right) \right] }{ a^2 \left( t^2 + 1 \right) }.
\end{equation}

The absolute value function seems somewhat tricky here, but we may consider conditions of either signs,
\begin{equation}\label{eq13}
    n_Z = n_W \cdot \frac{ \pm \left[ a^2 + 2atu + t^2 \left( u^2+v^2+w^2 \right) \right] }{ a^2 \left( t^2 + 1 \right) }.
\end{equation}

Applying the chain rule with~\eqref{eq09}, the Ricci scalar in physical space reduces to~\eqref{eq14} and the plus minus sign vanishes, which indicates that the invariance of Ricci curvature under 3D M\"obius mapping is proven.

\section{2D exponent conformal mappings}
2D exponent conformal mapping applied on all three ``1D'' gradient media to obtain the 2D versions respectively. We demonstrate the mathematical details in this section. The 2D conformal mapping reads
\begin{equation}
    W= e^{Z},
\end{equation}
where $Z=x+\mathrm{i}y$ is the coordinates in real space and $W=u+\mathrm{i}v$ is the coordinates in virtual space. Equivalently, we can expand the complex mapping into two real mappings,
\begin{equation}\label{eq15}
    u=e^{x}\cos y,\, v=e^{x} \sin y,
\end{equation}
revealing the essence of coordinate transformation. According to the conformal transformation optics~\cite{xuConformalTransformationOptics2015}, the refractive index profile in real space can be derived with
\begin{equation}
    n_{Z} = n_{W}\left| \frac{\mathrm{d}W}{\mathrm{d}Z} \right|,
\end{equation}
where $\left| \frac{\mathrm{d}W}{\mathrm{d}Z} \right|$ can be interpreted as either the norm of a complex differentiation or the determinant of the Jacobian matrix corresponding to the coordinate transformation from $\left\{ u,v \right\}$ to $\left\{ x,y \right\}$ due to the existence of Cauchy-Riemann conditions.

We can therefore obtain the refractive index profiles in ``1D'' version for the three cases. First, we start from the 2D Maxwell's fisheye lens in virtual space with gradient refractive index profile
\begin{equation}
    n_W=\frac{2}{1+u^2+v^2}.
\end{equation}
According to the conformal transformation optics~\eqref{eq15}, the ``1D'' version named the Mikaelian lens can be derived with
\begin{equation}
    n_{Z} = n_{W}e^{x}={1}/{\cosh x}.
\end{equation}

Then, we start from the 2D optical black hole in virtual space with a refractive index profile
\begin{equation}
    n_{W} = 1/\sqrt{u^2+v^2}.
\end{equation}
And the transformed lens in physical space yields
\begin{equation}
    n_{Z} = 1,
\end{equation}
which is simply a homogeneous lens.

Lastly, we set the Poincar\'e Disk in virtual space,
\begin{equation}
    n_W=\frac{2}{1-u^2-v^2}.
\end{equation}
And we can get the transformed lens in physical space as
\begin{equation}
    n_{Z}=1/\sinh x.
\end{equation}

\printbibliography